\documentstyle  {article}

\begin {document}
\noindent{\Large {\bf  Gravitational light bending in  Euclidean space}}

\bigskip
\hspace {4 mm} { \parbox {10 cm} {
\noindent  {  I E Bulyzhenkov }

{\small
\bigskip
\noindent {Institute of Spectroscopy RAS, Troitsk, M.o., 142092, Russia}

\bigskip
\noindent
Both the non-homogeneous slowness of electromagnetic waves
in gravitational fields and the frequency red shift
contribute to the gravitational light bending. This twofold
contribution explains the measured deflection of light rays
by the Sun under Euclidean geometry of space.

     }}

 \bigskip \bigskip

As known, the observed light deflection by the Sun is nearly twice
what Newton's mechanics predicts for particles [1,2].
The hypothetical analogy between particles and light waves
was employed by majority of authors
for speculations about the curved   three-space around
 sources of gravity.

However, there is no direct analogy in gravitation
between the mechanical particle and the electromagnetic wave.
The rest-mass particle keeps  the measured energy
$m{\sqrt {{g_{oo}}}/ {\sqrt {1 - (dl/d\tau)^2}} }$ along
its trajectory in a static gravitational field, for example,
while the measured energy
$\hbar  \omega  = k_o/{\sqrt {{g_{oo}}}}$ of the electromagnetic
wave changes along the light ray together with changes
of the proper time rate $d\tau = {\sqrt {{g_{oo}}}}dx^o$.
Moreover, the gravitational field works like a medium for
electromagnetic waves and leads to their slowness,
$n = {\sqrt {  {\tilde \varepsilon} {\tilde \mu}   } } =
{{{{g^{-1/2}_{oo}}} }},$ because  ${\bf D} = g_{oo}^{-1/2} {\bf E} $
and ${\bf B} = g_{oo}^{-1/2} {\bf H} $
in covariant the Maxwell equations [3].

General equations for light rays in static gravitational
fields ought to be derived from Fermat's principle [4],
\begin {equation}
\delta\!\int k_i dx^i = -\delta\!\int\!\hbar\omega n dl =
-k_o\delta\!\int {{dl}\over {{g_{oo} }} }
= -k_o\delta\!\int {  { \sqrt {
  du^2 + u^2d\varphi^2   }   }\over u^2 (1- GM u)^2   } = 0,
\end {equation}
where $n^2 k_o k^o = k_i k^i,  {{g_{oo}}}k^o
= k_o = const $,  $k_i = -nk_o g_{oo}^{-1/2}
dx^i/dl$,  ${\sqrt {g_{oo}}}= 1-GM u$,
$g_i=-g_{oi}/g_{oo}=0$,
$dl = {\sqrt { \delta_{ij} x^ix^j}}$  = $
{\sqrt {       dr^2 + r^2d\varphi^2}}$
($ r \equiv u^{-1} $, $\varphi$, and  $\vartheta = \pi/2$
are the spherical coordinates). It is worthy of note
that both the non-homogeneous wave slowness,
$n = n(g_{oo})  \neq const$, and the non-homogeneous
frequency, $\omega = \omega (g_{oo})  \neq const$ (red shift),
are responsible in (1) for the twofold curvature of light
rays in gravitational fields.

The variations of (1) with respect to  $ u $ and
 $ \varphi $
  leads to a couple
of ray-equations,
\begin {equation}
(1- GM u)^4 \left [ \left({du\over d\varphi}\right )^2 +
u^2    \right] = U_o^2 = const,
\end {equation}

\begin {equation}
{{d^2u}\over d\varphi^2} + u =  2U_o^2 {{GM }\over (1- GM u)^5}.
\end {equation}

A family of solutions,
 $u \equiv r^{-1}$ $ = r^{-1}_o sin \varphi $ + $2GM r^{-2}_o
(1 + cos \varphi)$ and $r_o^{-1} = U_o $, for both these
equations  might be found in weak fields,  if one ignores
all terms nonlinear in  $GMr_o^{-1}  \ll 1$. A propagation
of  light from $r(-\infty) = \infty , \varphi (-\infty)= \pi$
to  $r (+\infty)  \rightarrow  \infty , \varphi (+\infty)
 \rightarrow \varphi_{\infty} $ corresponds to the angular deflection
$ \varphi_{\infty} = arsin [-2GM r_o^{-1}(1 + cos\varphi_\infty)]$
$\approx -4GMr_o^{-1}$ from the initial light direction.
This result, derived under the flat three-interval $dl$,
coincides with   the measured   deflection, $-1,66''  \pm 0.18''$ [5],
of light rays by  the Sun  ($r_o = 6,96\times 10^5km$
and $4GM r_o^{-1} $ $=$ $1,75''$).

A four-interval equation for electromagnetic waves in non-stationary
gravitational fields may be derived
from the light wave-velocity, $dl/d\tau  = n^{-1}$,
in the non-dispersive medium or vacuum,
{\it i.e.}
\begin {equation}
n^2(g_{oo},g_i)\delta_{ij} dx^i dx^j = g_{oo}(dx^o - g_idx^i)^2,
\end {equation}
where the slowness $n(g_{oo}, g_i)$ is determined by the contribution
of gravity into the covariant Maxwell and wave equations.
This non-stationary slowness defines the wave and ray equations,
$k_\mu k^\mu = (1-n^2)(k_\mu V^\mu)^2$ and $\delta \int k_\mu dx^\mu = 0$,
for light in arbitrary gravitational fields [4].

 Note that Euclidean   metrics, $ \gamma_{ij}
\equiv g_{oi}g_{oj}g^{-1}_{oo} - g_{ij} = \delta_{ij} $,
is hold in (1) and (4) only for 3D space, rather than for 4D space-time.
All components of the metric tensor
$g_{\mu\nu}$ $ \neq diag $ $(+1, -1,-1,-1)$ are determined
by a particular distribution of external (for an observer)  matter.
In other words, the three-space with universal
Euclidean geometry may be  common for all observers,
while curved space-times are different pseudo-Riemannian
manifolds for different observers.

 Euclid's 3D geometry satisfies the
  observed  conservation of a system three-momentum
(and angular momentum) at all points of  three-space.
The homogeneous conservation law
for a sum of the three-vectors
is possible, in principle, for common three-spaces with
constant curvatures (positive, negative,
or zero). But only flat three-space, corresponded to all known experiments,
may open a simple way for coupling gravity and electromagnetism [6].

{\centerline  {\rule {80mm}{0.1mm}}}
\bigskip

\noindent [1] Einstein A 1911 Uber den Einfluss der Schwerkraft
auf die Ausbreitung
des Lightes {\it Annalen der Physik }  {\bf 35} 898;

\noindent [2] Lenard P 1921 Uber die Ablenkung eines
Lichtstrahls von seiner geradlinigen Bewegung
durch die Attraktion eines Weltk$\ddot  {\rm o}$rpers,
an welchem er nahe vorbeigeht, von J
Soldner 1801 {\it Annalen der Physik} {\bf 65} 593

 \noindent [3]  Landau L D and Lifshitz E M 1971
{\it The classical theory of fields} (Oxford: Pergamon)

\noindent [4]  Synge J L 1960 {\it Relativity: the General Theory}
(Amsterdam: North-Holland Publishing Company)

\noindent [5]  Will C M  1993 {\it Theory and experiment
in gravitational physics}
(Cambridge University Press)

\noindent [6] Bulyzhenkov I E 1997, e-preprint, gr-qc/9708004.
\end {document}